\documentclass[aps,twocolumn,superscriptaddress]{revtex4-1}
\usepackage[colorlinks=true,bookmarks=false,citecolor=magenta,urlcolor=magenta,linkcolor=magenta,breaklinks]{hyperref} 
\usepackage{breakurl}
\usepackage{sidecap}
\usepackage{amssymb}
\usepackage{hhline}
\sidecaptionvpos{figure}{t}
\usepackage{amsmath}
\usepackage{graphicx}
\usepackage{esint}
\usepackage{epstopdf}
\usepackage{rotating}
\graphicspath{{pict/}{./}}
\usepackage{bm}%
\usepackage{titlesec}

\newcounter{Fig}

\begin{document}

\title{Singularities and Poincar\'{e} Indexes of Electromagnetic Multipoles}
\author{Weijin Chen}
\affiliation{School of Optical and Electronic Information, Huazhong University of Science and Technology, Wuhan, Hubei 430074, P. R. China}
\author{Yuntian Chen}
\email{yuntian@hust.edu.cn}
\affiliation{School of Optical and Electronic Information, Huazhong University of Science and Technology, Wuhan, Hubei 430074, P. R. China}
\author{Wei Liu}
\email{wei.liu.pku@gmail.com}
\affiliation{College for Advanced Interdisciplinary Studies, National University of Defense
Technology, Changsha, Hunan 410073, P. R. China}
\begin{abstract}
Electromagnetic multipoles have been broadly adopted as a fundamental language throughout photonics, of which general features such as radiation patterns and polarization distributions are generically  known, while their singularities and topological properties have mostly slipped into oblivion. Here we map all the singularities of multipolar radiations of different orders, identify their indexes, and show explicitly the index sum over the entire momentum sphere is always $2$, consistent with the Poincar\'{e}-Hopf theorem. Upon those revealed properties, we attribute the formation of bound states in the continuum to the overlapping of multipolar singularities with open radiation channels. This insight unveils a subtle equivalence between indexes of multipolar singularities and topological charges of those bound states. Our work has fused two fundamental and sweeping concepts of multipoles and topologies, which can potentially bring unforeseen opportunities for many multipole related fields within and beyond photonics.
\end{abstract}
\maketitle
Electromagnetic multipoles characterized by vector spherical harmonics constitute a complete basis for light field expansions, and are playing indispensable roles in diverse branches
 of optics and photonics~\cite{jackson1962classical,Bohren1983_book,RAAB_2004__Multipole}. Those multipoles have been extensively studied and their general features, such as radiation patterns, polarization distributions, far-field parities and so on, have been comprehensively exploited for various applications. A rather outstanding example of this is the burgeoning field of meta-optics largely built on electric and magnetic multipoles of different orders, the interferences among which can render enormous extra freedom for manipulating light-matter interactions in both linear and nonlinear regimes~\cite{jahani_alldielectric_2016,KUZNETSOV_Science_optically_2016,SMIRNOVA_Optica_multipolar_2016,LIU_2018_Opt.Express_Generalized}.

In spite of the great achievements relying on electromagnetic multipoles, an unfortunate situation is that rare attention has been paid to those \textit{dark directions} along which there are no radiations. Those directions are naturally considered to be trivial; singularities and topological features of multipoles have neither attracted much interest nor properly investigated. This is rather puzzling, as from a mathematical viewpoint, in the far field multipolar radiations are transverse and perfectly makes an elementary case of tangent vector fields on a momentum sphere. Depending on the order of  multipoles, there are a finite number (at least one) of singularities (zeros or dislocations) of the vector fields. They are isolated directions where there are no radiations and for which  Poincar\'{e} indexes (winding numbers or topological invariants) can be assigned.  Moreover the Poincar\'{e}-Hopf theorem, or more precisely the hairy ball theorem for this specific case, can be simply applied~\cite{MILNOR_1997__Topology,RICHESON_2012__Euler,GBUR_2016__Singular,BERRY_2017__HalfCentury}. It is even more stunning that such a situation has been stagnant, withstanding the recent overwhelming trend of incorporating topological concepts into almost every branch of physics~\cite{Hasan2010_RMP,Qi2011_RMP,YANG_Phys.Rev.Lett._topological_2015-1,HUBER_2016_Nat.Phys._Topological,Lu2014_topological,OZAWA_2018_ArXiv180204173}. Considering that  electromagnetic multipoles broadly serve as a basic language for descriptions and explanations of various optical effects, to clarify their topological features is of apparently great significance, especially for the expanding field of topological photonics~\cite{Lu2014_topological,OZAWA_2018_ArXiv180204173}.

\begin{figure*}
  \begin{minipage}[c]{0.63\textwidth}
    \includegraphics[width=\textwidth]{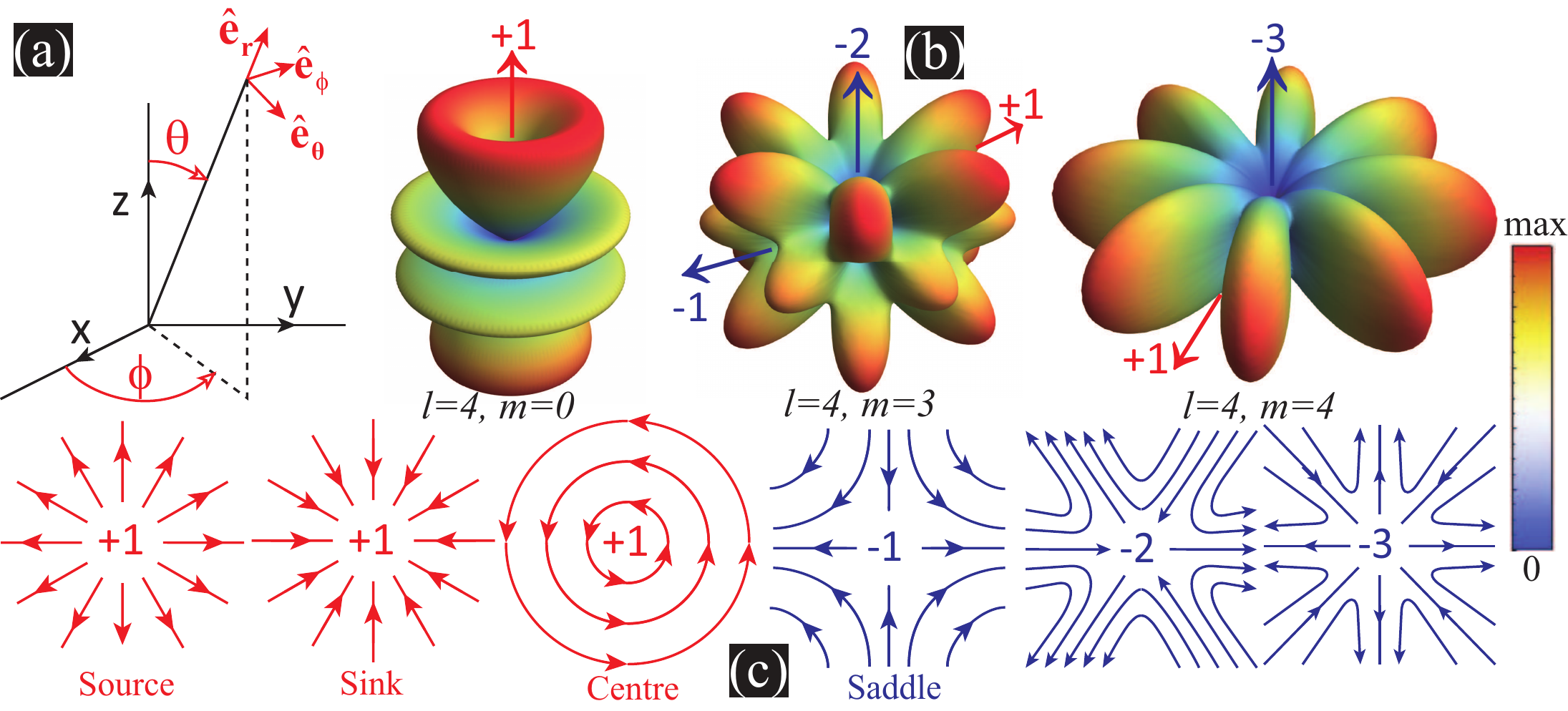}
  \end{minipage}\hfill
  \begin{minipage}[c]{0.32\textwidth}
    \caption{\small (a) Spherical polar coordinate system $(r, \theta, \phi)$  with the associated orthonormal basis vectors $\mathbf{\hat{e}}_\theta$, $\mathbf{\hat{e}}_\phi$ and $\mathbf{\hat{e}}_r$ indicated. (b) Radiation patterns for multipoles of order $(l, m)$. Singularities that represent each category of Table ~\ref{table} are pinpointed and the corresponding indexes are specified. (c) Vector field patterns close to singularities of different indexes from $-3$ to $+1$, which have covered all sorts of singularities indicated in (b).
    } \label{fig1}
  \end{minipage}
\end{figure*}

In this work we reapproach electromagnetic multipoles from a topological perspective, focusing on distributions of singularities and their indexes of the corresponding tangent vector fields on a $S^{2}$ momentum sphere. A comprehensive map is given, pinpointing all singularities with their indexes for multipoles of different orders. It is shown explicitly that of all multipoles the index sum over the entire momentum sphere is always $2$ (Euler characteristic of a sphere), consistent with the Poincar\'{e}-Hopf theorem~\cite{MILNOR_1997__Topology,RICHESON_2012__Euler}. Based on this revelation, we reinterpret the recently observed  bound states in the continuum (BICs)~\cite{HSU_Nat.Rev.Mater._bound_2016} from a multipolar perspective, and trace the mechanism to overlapping radiation singularities with open radiation channels. We further discover that topological charges of the BICs demonstrated have a multipolar underlying structure, which essentially are equivalent to singularity indexes of corresponding multipolar radiations. The incorporation of topological terminology into classical multipoles can potentially rejuvenate many multipole related studies, paving the way for introducing established topological concepts and advanced functionalities into more disciplines both within and beyond photonics~\cite{GBUR_2016__Singular,BERRY_2017__HalfCentury}.

Electromagnetic multipoles can be categorized into magnetic and electric groups, the radiated electric (magnetic) fields of which are characterized by vector spherical harmonics $\textbf{M}_{lm}$ ($\textbf{N}_{lm}$)  and $\textbf{N}_{lm}$ ($\textbf{M}_{lm}$), respectively~\cite{jackson1962classical,Bohren1983_book,Supplemental_Material}.  For analysis convenience one specific expression of $\textbf{M}_{lm}$ is chosen:~\cite{Bohren1983_book,Supplemental_Material}:
\begin{equation}
\label{tm_even}
\begin{aligned}
{{\bf{M}}_{lm}} = {\rm{ - }}{{\rm{m}} \over {\sin \theta }}\sin (m\phi )P_l^m\left( {\cos \theta } \right){\mathbf{\hat{e}}_\theta }\\
 - \cos (m\phi ){{dP_l^m\left( {\cos \theta } \right)} \over {d\theta }}{\mathbf{\hat{e}}_\phi}.
\end{aligned}
\end{equation}
Here we have dropped the radially dependent terms to focus on the tangent fields; $P_l^m\left( {\cos \theta } \right)$  ($-l\leq m \leq l$) denotes associated Legendre polynomials~\cite{BRONSHTEIN_2007__Handbook}; basis vectors $\mathbf{\hat{e}}_\theta$, $\mathbf{\hat{e}}_\phi$ and $\mathbf{\hat{e}}_r$, polar and azimuthal angles $\theta$ and $\phi$ are shown in Fig.~\ref{fig1}(a); $\textbf{M}_{lm}\cdot\mathbf{\hat{e}_r}=0$ and $\textbf{N}_{lm}\cdot\textbf{M}_{lm}=0$. As a first step, we investigate the poles with $|\cos \theta|=1$, where $\phi$ is not defined. Except for $|m|=1$, both are singularities and the index is~\cite{Supplemental_Material}:
\begin{equation}
\label{index_ns}
{\mathbf{Ind}}=1-|m|,~~{|\cos \theta|=1}.
\end{equation}
When $m=\pm 1$, the poles are not singular, for which we can still assign the index of $0$ and thus Eq.~(\ref{index_ns}) is still applicable~\cite{Supplemental_Material}.

Next we turn to regions excluding the poles, where for $m=0$ there are no singularities~\cite{Supplemental_Material}. When $m\neq0$, according to Eq.~(\ref{tm_even}), a singularity requires that both components along $\mathbf{\hat{e}}_\theta$ and $\mathbf{\hat{e}}_\phi$ are zero, and thus indexes and positions of singularities are~\cite{Supplemental_Material}:
\begin{equation}
\label{index_other}
\begin{aligned}
{\mathbf{Ind}}=-1,~~\cos (m\phi)=P_l^m\left( {\cos \theta } \right)=0;\\
{\mathbf{Ind}}=+1,~~\sin (m\phi)=\frac{dP_l^m({\cos \theta })}{d\theta}=0.
\end{aligned}
\end{equation}
For $0<\theta<\pi$ and $0\leq \phi \leq 2\pi$: both $\sin (m\phi )=0$ and $\cos (m\phi )=0$ have $2|m|$ solutions; while equations of $P_l^m\left( {\cos \theta } \right)=0$ and $\frac{dP_l^m({\cos \theta })}{d\theta}=0$ have $l-|m|$ and $l-|m|+1$ solutions, respectively~\cite{BRONSHTEIN_2007__Handbook}.  It means that except for the poles, there are
$2|m|(l-|m|)$ singularities with index of $-1$ and $2|m|(l-|m|+1)$ singularities with index of $+1$ (this is still a valid statement for $m=0$~\cite{Supplemental_Material}).  Basically the index sum for all singularities is $2(1-|m|)+2|m|(l-|m|+1)-2|m|(l-|m|)$, which would always make up to $2$ and agrees with the Poincar\'{e}-Hopf theorem~\cite{MILNOR_1997__Topology,RICHESON_2012__Euler}.

Up to now, we have discussed only magnetic multipoles characterized by ${{\bf{M}}_{lm}}$ with one specific expression shown in Eq.~(\ref{tm_even}).  For the other expression of ${{\bf{M}}_{lm}}$ [with $\sin (m\phi )$ and $\cos (m\phi )$ interchanged and $\mathbf{\hat{e}}_\theta$ changed to $-\mathbf{\hat{e}}_\theta$~\cite{Bohren1983_book,Supplemental_Material}), identical distribution map for singularities and indexes would be obtained except for an interchange between $\sin (m\phi )$ and $\cos (m\phi )$ in Eq.~(\ref{index_other}). For electric multipoles characterized by ${{\bf{N}}_{lm}}$, the duality of Maxwell equations requires that the corresponding tangent fields can be obtained directly from those of ${{\bf{M}}_{lm}}$ (the terms with radial dependence are dropped for  studies of tangent fields) by the following transformation~\cite{Bohren1983_book,Supplemental_Material}: $\mathbf{\hat{e}}_\theta \rightarrow \mathbf{\hat{e}}_\phi$ and $\mathbf{\hat{e}}_\phi \rightarrow -\mathbf{\hat{e}}_\theta$. Since the transformation above would neither change the position nor the index of the singularity~\cite{Supplemental_Material}, as a result both electric and magnetic multipoles of the same order $(l, m)$ share the same singularity and index distributions, as summarized in Table ~\ref{table}. For example,  Fig.~\ref{fig1}(b)  shows radiation patterns and pinpoint some representative singularities with indexes for three multipoles (see Ref.~\cite{Supplemental_Material} for more scenarios). The vector field patterns close to singularities of different indexes are shown in Fig.~\ref{fig1}(c). For the case of index $+1$, the singular point could be a source (sink) or a centre, depending on which field (electric or magnetic) we employ to show the field lines~\cite{Supplemental_Material}.

\begin{table}[htp]
\centering 
\begin{tabular}{| c | c | c | c|} 
\hline
Positions & Index & Number & Total Indexes \\ [0.5ex] 
\hline 
$|\cos(\theta)|=1$ & $1-|m|$ & 2 & $2(1-|m|)$ \\ 
\hline
$\cos (m\phi )=P_l^m\left( {\cos \theta } \right)$ & & $2|m|\times$ & $-2|m|\times$\\$=0,|\cos(\theta)|\neq1$  &\raisebox{1.5ex}{-1 $(m\neq0)$} &  \raisebox{0ex}{$(l-|m|)$} & \raisebox{0ex}{$(l-|m|)$} \\
\hline 
\raisebox{-1.3ex}{$\sin (m\phi )=\frac{dP_l^m({\cos \theta })}{d\theta}$} & \raisebox{-0.75ex} {0 $(m=0)$} & \raisebox{-0.75ex} {$2|m|(l-$} & \raisebox{-0.75ex}{$2|m|\times$}\\$=0,|\cos(\theta)|\neq1$  & \raisebox{0.75ex} {+1 $(m\neq0)$} &  \raisebox{0.75ex}{$|m|+1)$} & \raisebox{0.75ex}{$(l-|m|+1)$} \\
\hline
\end{tabular}\\
\caption{Distributions of the singularities and their indexes for radiated vector fields from a multipole of order $(l, m)$. } 
\label{table} 
\end{table}

Table~\ref{table} tells that a singularity with index larger than $+1$ is not accessible with an individual multipole. In contrast, if a series of multipoles of different orders and/or natures are combined together, a singularity with larger positive index is always accessible, which can be simply justified by the following arguments: for a specific index, we can always predesign a continuous tangent vector field that includes such a singularity; the electromagnetic multipoles constitute an orthogonal and complete bases for vector field expansion; the predesigned vector field can always be expanded into a set of multipoles, with expansion coefficients of $a_{lm}$ and $b_{lm}$ that are associated with $\textbf{N}_{lm}$ and $\textbf{M}_{lm}$, respectively~\cite{Supplemental_Material}.

\begin{figure}[tp]
\centerline{\includegraphics[width=8.5cm]{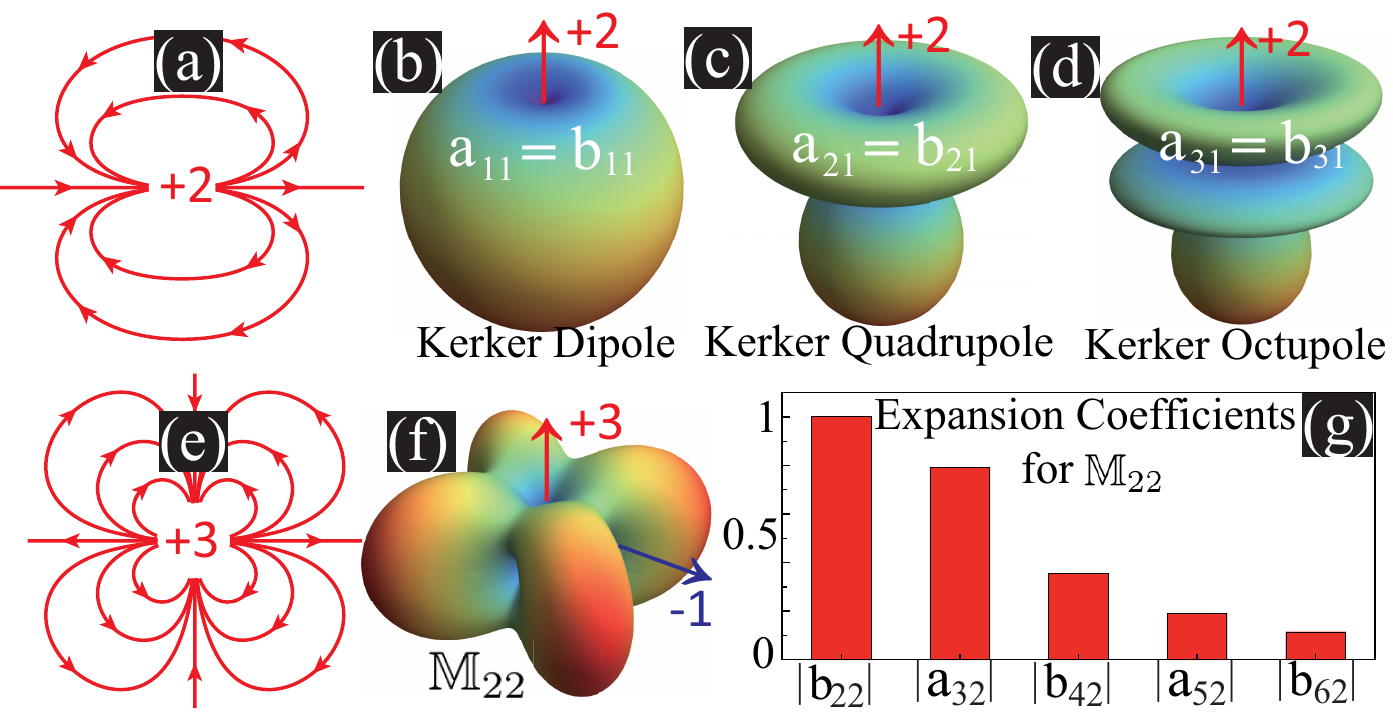}} \caption{\small (a) and (e): Vector field patterns close to singularities of indexes $+2$ and $+3$. (b)-(d) Multipolar combinations (Kerker dipole, quadrupole and octupole) that end up with only one singularity of index $+2$. (f) Radiation patterns and singularity distributions for transformed $\mathbb{M}_{22}$, which has to be expanded into a series of multipoles, with normalized magnitudes of expansion coefficients (up to $l=6$) shown in (g).}
\label{fig2}
\end{figure}

According to the Poincar\'{e}-Hopf theorem~\cite{MILNOR_1997__Topology,RICHESON_2012__Euler}, the index sum for all singularities over the momentum sphere should always be $2$, being the radiations from an isolated multipole or a set of multipoles. The simplest allowed case of this theorem is that there is only one singularity (at least one singularity) and the index thus has to be $+2$. This corresponds to (generalized) Kerker conditions (Kerker multipoles) or (generalized) Huygens sources~\cite{jahani_alldielectric_2016,KUZNETSOV_Science_optically_2016,LIU_2018_Opt.Express_Generalized}. The field pattern close to a singularity of index $+2$ (also termed as dipole singularity~\cite{MILNOR_1997__Topology,RICHESON_2012__Euler}) is shown in Fig.~\ref{fig2}(a).  This type of singularity can be obtained from interferences of a pair of electric and magnetic multipoles of the same order, magnitude and phase ($a_{lm}=b_{lm}$, $m=\pm1$), but opposite parities ~\cite{Liu2014_ultradirectional,LIU_Phys.Rev.Lett._generalized_2017}, with the first three cases of overlapping dipoles, quadrupoles, and octupoles shown in Figs.~\ref{fig2}(b)-(d).

For a tangent and continuous vector field $\mathbb{M}_{lm}$ obtained from Eq.~(\ref{tm_even}) by forcing the following transformation (forbidden for $m=\pm1$ that breaks the continuity of the vector field~\cite{Supplemental_Material}) $\mathbf{\hat{e}}_\theta \rightarrow -\mathbf{\hat{e}}_\theta$ (or equivalently $\mathbf{\hat{e}}_\phi \rightarrow -\mathbf{\hat{e}}_\phi$):
\begin{equation}
\label{tm_transformed}
\begin{aligned}
{{\mathbb{M}}_{lm}} = {\rm{}}{{\rm{m}} \over {\sin \theta }}\sin (m\phi )P_l^m\left( {\cos \theta } \right){\mathbf{\hat{e}}_\theta }\\
 - \cos (m\phi ){{dP_l^m\left( {\cos \theta } \right)} \over {d\theta }}{\mathbf{\hat{e}}_\phi}.
\end{aligned}
\end{equation}
The singularity distribution table for $\mathbb{M}_{lm}(|m|\neq 1)$ would be the same as Table~\ref{table}, except that the index in the second column is changed from $1-|m|,~-1,~+1$ to $1+|m|,~+1,~-1$~\cite{Supplemental_Material}. This means that for such a transformed vector field, the index for both singularities at poles would be $1+|m|$. For example, in Fig.~\ref{fig2}(e) we show the field patterns close to a singularity of index $+3$, which is present on the poles of the radiation pattern of  $\mathbb{M}_{22}$ (shown in Fig.~\ref{fig2}(f); despite the index difference, the radiation pattern is the same as that of $\textbf{M}_{22}$~\cite{Supplemental_Material}). As has been already explained, $\mathbb{M}_{22}$ cannot be represented by an individual multipole, but by a series of multipoles, with the multipolar composition (normalized magnitudes of expansion coefficients) shown in Fig.~\ref{fig2}(g).

After mapping out the singularities and their indexes for electromagnetic multipoles, now we proceed to show how those properties could be powerfully useful, taking recently demonstrated BICs as an example~\cite{HSU_Nat.Rev.Mater._bound_2016}. For convenience of multipolar analysis, the periodic structure can be approached from an alternative reductionist perspective~\cite{LEPETIT_2014_Phys.Rev.B_Controlling,MONTICONE_2017_NewJ.Phys._Bound,LIU_2017_ACSPhotonics_Beam,DOELEMAN_2018_Nat.Photonics_Experimentala,HA_2018_Nat.Nanotechnol._Directionala}: it can be treated as an infinite ensemble of radiating items (unit cells); the overall optical properties of the periodic structure can be interpreted as interferences of radiations from all the unit cells. For an isolated radiating item, there are an infinite number of open out-coupling radiation channels, corresponding to all the points (directions) on the momentum sphere. While for the periodic structure, there are only a finite number of such channels, corresponding to diffractions channels of different orders along certain directions. From this perspective, the formation of a BIC can be attributed to overlapping of the radiation singularities of each unit cell with the open diffraction channels of the periodic structure. An index could be assigned to the singularity of the unit-cell radiation that overlaps with the open channel (there can be extra singularities not overlapping with such channels), which is exactly the topological charge of the induced BIC~\cite{HSU_Nat.Rev.Mater._bound_2016,ZHEN_2014_Phys.Rev.Lett._Topological}.

\begin{figure}[tp]
\centerline{\includegraphics[width=9cm]{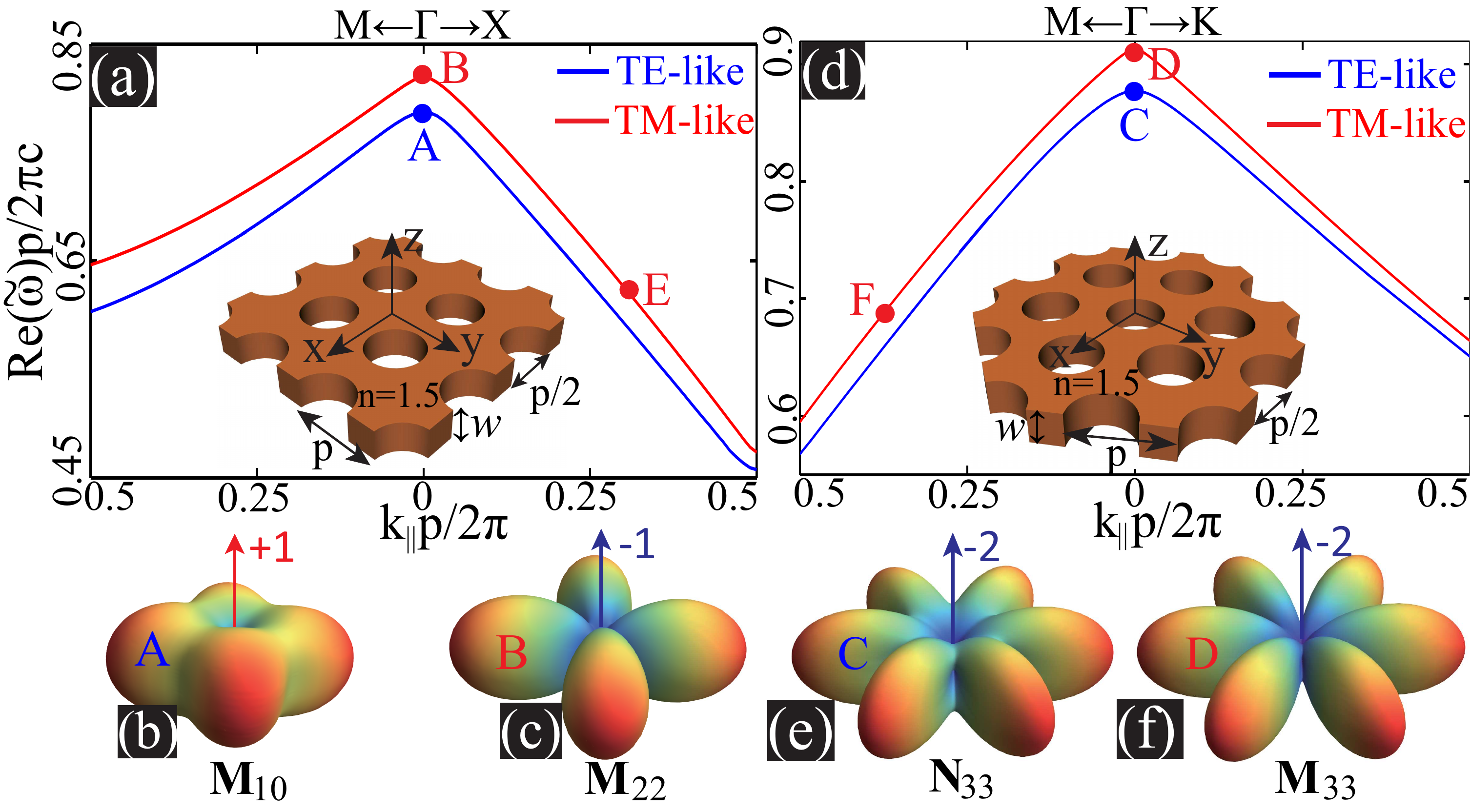}} \caption{\small (a) and (d): Dispersion curves for square and hexagonal photonic crystal slabs and six BICs are indicated by points \textbf{A}-\textbf{F}: ${{\rm{k}}_{\parallel}} =\sqrt{k_x^2 + k_y^2}$ is on-plane angular wavenumber and $c$ is the speed of light. The corresponding radiation patterns and the dominant multipolar components of the four $\Gamma$-point BICs are shown in (b)-(c) and (e)-(f).}\label{fig3}
\end{figure}

To further crystallize what has been stated above, we turn to the photonic crystal slabs of square or hexagonal lattices, which are wieldy employed for investigations into BICs~\cite{HSU_Nat.Rev.Mater._bound_2016,YANG_2014_Phys.Rev.Lett._Analytical,ZHEN_2014_Phys.Rev.Lett._Topological,HSU_Nature_observation_2013-1,GUO_Phys.Rev.Lett._topologically_2017,KODIGALA_Nature_lasing_2017,ZHANG_2018_Phys.Rev.Lett._Observation}.
For a direct comparison between our results from multipolar analysis with those established ones, and to connect the singularity indexes of multipolar radiations to the confirmed topological charges of BICs, we investigate structures similar to those studied in Refs.~\cite{ZHEN_2014_Phys.Rev.Lett._Topological,HSU_Nature_observation_2013-1}. The photonic crystal slabs of square or hexagonal lattices of circular air holes (slab refractive index $n=1.5$; width $w$; lattice constant $p$; air hole diameter $p/2$) are studied, which are shown as insets in Figs.~\ref{fig3}(a) and (d). For both cases, we show the dispersion curves of two representative bands (one TE-like and one TM-like bands;  $\widetilde{\omega}$ is the complex mode eigenfrequency~\cite{Supplemental_Material}).

Firstly we check the symmetry-protected BICs located on the $\Gamma$-points. For both lattices, there are two such BICs on both the TE-like and TM-like bands [four BIC points \textbf{A}-\textbf{D} are indicated in Figs.~\ref{fig3}(a) and (d), with $w=p/2$]. For each of them, we show in Figs.~\ref{fig3}(b)-(c) and (e)-(f) the  corresponding major dominant multipolar components (see Ref.~\cite{Supplemental_Material} for details of multipolar expansions based on near-field currents) and the far-field radiation patters. The four symmetry-protected BICs have a common feature that unit-cell radiation corresponds to that of a dominant individual multipole with fixed order and the radiation singularity overlaps with the allowed radiation channel (radiation is zero along $\mathbf{z}$ direction). This agrees with our previous discussions with regard to the multipolar interpretation of the underlying mechanism of BICs. Moreover, based on the dominant multipolar components, for each singularity along $\mathbf{z}$ direction an index can be directly assigned according to the $|\cos(\theta)|=1$ category in Table~\ref{table} (other minor multipoles not shown modifies a bit the overall radiation patterns, while not affecting singularity and index distributions~\cite{Supplemental_Material}), which is exactly the topological charge revealed in Ref.~\cite{ZHEN_2014_Phys.Rev.Lett._Topological}.

\begin{figure}[tp]
\centerline{\includegraphics[width=8.5cm]{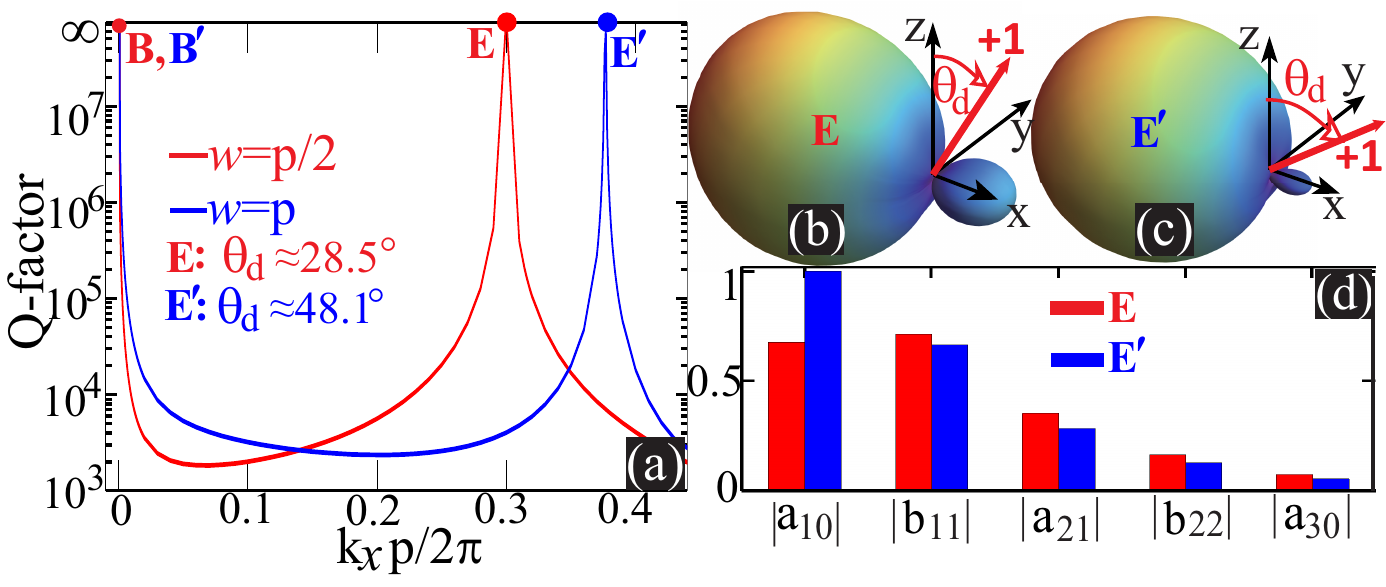}} \caption{\small (a) With width $w$ changed from $p/2$ to $p$, the off-$\Gamma$-point BIC is shifted, while the $\Gamma$-point BIC  is fixed. The multipolar compositions for the two off-$\Gamma$-point BICs are shown in (d) (only the major contributing terms are included), and the reconstructed far-field radiation patterns are shown in (b) and (c). For both cases the singularities are of index $+1$  and overlap with the open diffraction channels.}
\label{fig4}
\end{figure}

Now we turn to the off-$\Gamma$-point accidental BICs~\cite{ZHEN_2014_Phys.Rev.Lett._Topological,HSU_Nature_observation_2013-1}, taking the indicated point $\textbf{E}$ ($k_xp/2\pi=0.299$) in Fig.~\ref{fig3}(a) on the TM-like band for example [see similar analysis in Ref.~\cite{Supplemental_Material} for indicated BIC point $\textbf{F}$ in Fig.~\ref{fig3}(d) of the hexagonal lattice]. We show respectively both its multipolar composition and reconstructed far-field patterns of the unit cell in Figs.~\ref{fig4}(d) and (b). As expected, Fig.~\ref{fig4}(b) shows that a pair of singularities (both of index $+1$ guaranteed by $\sigma_z$ mirror symmetry) overlap with the allowed diffraction directions [$\phi_d=0$ and $\theta_d=\arcsin(k_x/k_0)$, where $k_0$ is total the angular wavenumber]. Moreover, the index $+1$ of the singularity agrees with the topological charge previously revealed~\cite{ZHEN_2014_Phys.Rev.Lett._Topological}.  This BIC would move to another position $\textbf{E$^\prime$}$ ($k_xp/2\pi=0.377$) by changing $w$, as is shown in Fig.~\ref{fig4}(a). It is clear from Figs.~\ref{fig4}(c) and (d) that the multipolar composition changes accordingly such that the multipolar singularity pair coincides with the new diffraction direction, which ensures that the BIC can move smoothly to the new position of different $k_x$ (see Ref.~\cite{Supplemental_Material} for decay states which correspond to separated singularities and open channels). 

Also according to Fig.~\ref{fig4}(a), the symmetry-protected BIC is fixed at $\Gamma$-point with changing $w$, which can certainly be explained through symmetry analysis~\cite{ZHEN_2014_Phys.Rev.Lett._Topological,KODIGALA_Nature_lasing_2017,ZHANG_2018_Phys.Rev.Lett._Observation,SAKODA_2004__Optical}. Our multipolar interpretation gives a different insight: for this type of BIC, the unit-cell radiation is represented by multipoles of order $|m|\neq 1$~\cite{Supplemental_Material}, the radiations of which are intrinsically zero along $\mathbf{z}$ direction (see Figs.~\ref{fig1} and ~\ref{fig3}, and Ref.~\cite{Supplemental_Material}) and can thus maintain the BIC on $\Gamma$-point. In sharp contrast, for the accidental BICs shown in Fig.~\ref{fig4}, zero radiation is accidental that originates from completely destructive multipolar interferences. Changing geometric parameters would alter the multipolar ratios and thus also the singularity positions, inevitably shifting the BICs. Moreover, as we have argued before, if the unit-cell radiation is represented by a dominant individual multipole, it is not possible to obtain positive index (topological charge) larger than $+1$ (see Table ~\ref{table}), no matter which band we rely on. Though symmetry analysis reveals that large positive index is not forbidden ~\cite{ZHEN_2014_Phys.Rev.Lett._Topological,KODIGALA_Nature_lasing_2017,ZHANG_2018_Phys.Rev.Lett._Observation,SAKODA_2004__Optical}, here we make a further step to show more microscopically that, to obtain a larger positive index, at least two combined multipoles are required (see Fig.~\ref{fig2}). This is probably synonymous with the phenomenon that: in previous studies large negative-index BIC have been widely achieved~\cite{ZHEN_2014_Phys.Rev.Lett._Topological,KODIGALA_Nature_lasing_2017,ZHANG_2018_Phys.Rev.Lett._Observation,SAKODA_2004__Optical}, whereas their positive counterparts rarely manifest themselves.

In conclusion, we revisit electromagnetic multipoles from a topological perspective, and provide a comprehensive map for the distributions of singularities and their indexes for multipolar radiations. It's shown that for an individual multipole, it is not possible to obtain singularity index larger than +1; while for the combination of a series of multipoles there is no limit for its singularity, but only requires that index sum of all singularities over the momentum space has to be $2$. Singularities of multipolar radiations are synonymous with the formation of BICs, as long as they coincide with open radiation channels. Based on this multipolar revelation, we further uncover the subtle equivalence between singularity indexes of multipoles and topological charges of BICs.  Considering the ubiquitous roles of electromagnetic multipoles all across photonics, our work will accelerate the pervasion of topological concepts into more optical branches, bring unperceived opportunities for various applications. Furthermore, multipoles basically serve as a fundamental tool and language for many other fields involving wave effects, on which our work generally shed new light from a fundamental topological perspective.

We acknowledge the financial support from National Natural Science Foundation of China (Grant No. 11874026, 11404403 and 11874426), and the Outstanding Young Researcher Scheme of National University of Defense Technology.

\setcounter{figure}{0}
\setcounter{table}{0}
\setcounter{equation}{0}

\makeatletter
\renewcommand{\thefigure}{S\@arabic\c@figure}
\makeatother

\makeatletter
\renewcommand{\theequation}{S\@arabic\c@equation}
\makeatother

\makeatletter
\renewcommand{\thetable}{S\@arabic\c@table}
\makeatother

\textbf{\textsc{Supplemental Material for ``Singularities and Poincar\'{e} Indexes of Electromagnetic Multipoles"}}\\

\textit{This Supplemental Material includes the following sections: \textbf{\uppercase\expandafter{\romannumeral1}}. Expressions for vector spherical harmonics; \textbf{\uppercase\expandafter{\romannumeral2}}. Positions and indexes for singularities of the multipolar radiations; \textbf{\uppercase\expandafter{\romannumeral3}}. Radiation patterns, singularities and indexes for $\mathbf{M}^e_{lm}$ with $l=1\rightarrow3$ and $m=0\rightarrow l$; \textbf{\uppercase\expandafter{\romannumeral4}}. Multipolar expansions and the radiation patterns for unit cells of infinite periodic structures; \textbf{\uppercase\expandafter{\romannumeral5}}. Numerical calculations of the dispersion bands, the mode Q-factors and the field distributions within the unit cells of photonic crystal slabs; \textbf{\uppercase\expandafter{\romannumeral6}}. Decay (non-BIC) states and separations of multipolar singularities from open radiation channels; \textbf{\uppercase\expandafter{\romannumeral7}}. Multipolar compositions for the $\Gamma$-point BICs; \textbf{\uppercase\expandafter{\romannumeral8}}. Off-$\Gamma$-point BICs of the hexagonal photonic crystal slab.}

\titleformat{\section}[hang]{\bfseries}{\thesection.\ }{0pt}{}
\renewcommand{\thesection}{\Roman{section}}
\renewcommand{\thesubsection}{\thesection.\Roman{subsection}}

\section{Expressions for vector spherical harmonics}

Vector spherical harmonics can be generated from a scalar function $\Psi$ through the following relations~\cite{Bohren1983_book}:
\begin{equation}
\label{generation}
\mathbf{M}=\bigtriangledown \times (\mathbf{r}\Psi),~~ \mathbf{N}=\frac{\bigtriangledown \times \mathbf{M}}{k},
\end{equation}
where $\mathbf{r}$ is the radial vector and $k$ is angular wavenumber. $\Psi$ can be an even or odd function with respect to azimuthal angle $\phi$, which are respectively expressed as~\cite{Bohren1983_book}:
\begin{equation}
\label{scalar}
\begin{split}
\Psi^{e}_{lm}=\cos(m\phi)P_l^m\left( {\cos \theta } \right)z_n(kr),\\
\Psi^{o}_{lm}=\sin(m\phi)P_l^m\left( {\cos \theta } \right)z_n(kr),
\end{split}
\end{equation}
where $P_l^m\left( {\cos \theta } \right)$  ($-l\leq m \leq l$) are associated Legendre polynomials and $z_n(kr)$ is the spherical Bessel or Hankel function~\cite{BRONSHTEIN_2007__Handbook}. Eqs.~(\ref{generation}) and (\ref{scalar}) would lead to the two sets of expressions for both $\mathbf{M}$ and $\mathbf{N}$~\cite{Bohren1983_book} (in this work as we fucus on the radiated tangent vector fields, the $\mathbf{r}$-dependent terms are dropped and all components along $\mathbf{r}$ are neglected):
\begin{equation}
\label{harmonics}
\begin{split}
\begin{aligned}
{{\bf{M}}^e_{lm}} = {\rm{ - }}{{\rm{m}} \over {\sin \theta }}\sin (m\phi )P_l^m\left( {\cos \theta } \right){\mathbf{\hat{e}}_\theta }\\
 - \cos (m\phi ){{dP_l^m\left( {\cos \theta } \right)} \over {d\theta }}{\mathbf{\hat{e}}_\phi},
\end{aligned}\\
\begin{aligned}
{{\bf{M}}^o_{lm}} = {\rm{}}{{\rm{m}} \over {\sin \theta }}\cos (m\phi )P_l^m\left( {\cos \theta } \right){\mathbf{\hat{e}}_\theta }\\
 - \sin (m\phi ){{dP_l^m\left( {\cos \theta } \right)} \over {d\theta }}{\mathbf{\hat{e}}_\phi},
\end{aligned}\\
\begin{aligned}
{{\bf{N}}^e_{lm}} = {\rm{ - }}{{\rm{m}} \over {\sin \theta }}\sin (m\phi )P_l^m\left( {\cos \theta } \right){\mathbf{\hat{e}}_\phi }\\
 + \cos (m\phi ){{dP_l^m\left( {\cos \theta } \right)} \over {d\theta }}{\mathbf{\hat{e}}_\theta},
\end{aligned}\\
\begin{aligned}
{{\bf{N}}^o_{lm}} = {\rm{}}{{\rm{m}} \over {\sin \theta }}\cos (m\phi )P_l^m\left( {\cos \theta } \right){\mathbf{\hat{e}}_\phi }\\
 + \sin (m\phi ){{dP_l^m\left( {\cos \theta } \right)} \over {d\theta }}{\mathbf{\hat{e}}_\theta}.
\end{aligned}\\
\end{split}
\end{equation}
It is clear from Eq.~(\ref{harmonics}) that we can convert ${{\bf{M}}^{e,o}_{lm}}$ directly to ${{\bf{N}}^{e,o}_{lm}}$ through the following transformation (duality transformation): $\mathbf{\hat{e}}_\theta \rightarrow \mathbf{\hat{e}}_\phi$ and $\mathbf{\hat{e}}_\phi \rightarrow -\mathbf{\hat{e}}_\theta$, which is guaranteed by the duality of Maxwell equations~\cite{jackson1962classical}. In the main letter, we have chosen ${{\bf{M}}^{e}_{lm}}$ for analysis. The analysis shown below about singularity and index distributions can be easily extended to ${{\bf{M}}^{o}_{lm}}$ and ${{\bf{N}}^{e,o}_{lm}}$, and the same conclusions can be drawn.

\section{Positions and indexes for singularities of the multipolar radiations}

For the multipole characterized by ${{\bf{M}}^{e}_{lm}}$, the singularity requires that£º
\begin{equation}
\label{singular}
|{{\bf{M}}^{e}_{lm}}|=0.
\end{equation}
The Poincar\'{e} index of the singularity located at $(\theta_s,\phi_s)$ can be defined as~\cite{RUDIN}:
\begin{equation}
\label{index}
{\rm{\mathbf{Ind} (\theta_s,\phi_s)= }}{1 \over {2\pi }}\oint_\mathbf{\gamma} {d\alpha },
\end{equation}
where $\mathbf{\gamma}$ is a continuous differentiable closed contour containing this singularity (all other singularities locate outside of the contour) and $d\alpha$ is the rotation angle of the vector field with respect to ${\mathbf{\hat{e}}_r }(\theta_s,\phi_s)$. Basically when the index is positive (negative), the rotation of the field vector is in the same (opposite) direction that the contour is traversed.

\subsection{Singularity and index at poles of $\theta=0$ or $\pi$}

$\blacksquare$~~~\textsc{{For the case of $m\neq \pm1$}}\\

It is easy to see that $|{{\bf{M}}^{e}_{lm}}|=0$ when $|m|\neq1$ and $\theta=0$ or $\pi$~\cite{BRONSHTEIN_2007__Handbook}, which means that multipoles with $|m|\neq1$ would have two singularities located on the two poles. For $\theta=0$ or $\pi$, the total rotation angle of the vector field close to singularity (along $\mathbf{z}$ or $-\mathbf{z}$) can be expressed as (here we assume that the contour is traversed in a counter-clockwise manner):
\begin{equation}
\label{angle}
\oint_\mathbf{\gamma(\theta\rightarrow0,\pi)} {d\alpha}=2\pi+2\pi|m|\sigma,
\end{equation}
where $\sigma$ is the helicity of the vector field in the rotating  orthonormal frame of $({\mathbf{\hat{e}}_\theta},~{\mathbf{\hat{e}}_\phi},~{\mathbf{\hat{e}}_r})$. Similar to the definition of circularly polarized light~\cite{Yariv2006_book_photonics}: $\sigma=1$ when the phase lag between the ${\mathbf{\hat{e}}_\theta}$ and ${\mathbf{\hat{e}}_\phi}$ components is $\beta_{{\mathbf{\hat{e}}_\phi}}-\beta_{{\mathbf{\hat{e}}_\theta}}=-\pi/2$ and thus the vector field is rotating in a counter-clockwise manner in the rotating frame; or  $\sigma=-1$  when $\beta_{{\mathbf{\hat{e}}_\phi}}-\beta_{{\mathbf{\hat{e}}_\theta}}=\pi/2$ and the vector field is rotating in a clockwise manner in the rotating frame; an extra special case is that $m=0$,  which means that there is only ${{\mathbf{\hat{e}}_\phi}}$ or  ${{\mathbf{\hat{e}}_\theta}}$ component [see Eq.~(\ref{harmonics})], indicating a \textit{linear polarization} and thus $\sigma=0$. Now the origin of the two terms on the right-hand side of Eq.~(\ref{angle}) is clear: the term $2\pi$ comes from the rotation of the frame itself; while the term $2\pi|m|\sigma$ comes from the ration of the vector field within the rotating frame.\\

For all the vector fields expressed by Eq.~(\ref{harmonics}), it is easy to confirm that $\sigma=0$ for $m=0$ and $\sigma=-1$ for $|m|\neq0$. From Eqs.~(\ref{index}) and (\ref{angle}) we obtain:
\begin{equation}
\label{index_pole}
{\rm{\mathbf{Ind} (\theta=0,\pi)= }}1-|m|,
\end{equation}
which is valid for $m=0$. Equation~(\ref{index_pole}) confirms that for multipoles characterized by both ${{\bf{M}}_{lm}}$ and ${{\bf{N}}_{lm}}$, the singularity index at the poles are identical, confirming that the duality transformation would neither change the position nor the index of the singularity.\\

$\blacksquare$~~~\textsc{{For the case of $m=\pm 1$}}\\

When $m=\pm 1$,  $|{{\mathbf{M}}^{e}_{lm}}|\neq0$ for $\theta=0$ or $\pi$. This means that the vector field is not singular at the south or north poles, for which we can assign an index of $0$, and thus  Eq.~(\ref{index_pole}) is still valid.\\

$\blacksquare$~~~\textsc{{Singularity and index distribution for vector fields of $\mathbb{M}^e_{lm}$} at $\theta=0$ or $\pi$}.\\

For a tangent and continuous vector field $\mathbb{M}^e_{lm}$ obtained from $\mathbf{M}^e_{lm}$ by forcing the following transformation  $\mathbf{\hat{e}}_\theta \rightarrow -\mathbf{\hat{e}}_\theta$ (or equivalently $\mathbf{\hat{e}}_\phi \rightarrow -\mathbf{\hat{e}}_\phi$):
\begin{equation}
\label{tm_transformed}
\begin{aligned}
{{\mathbb{M}}^e_{lm}} = {\rm{}}{{\rm{m}} \over {\sin \theta }}\sin (m\phi )P_l^m\left( {\cos \theta } \right){\mathbf{\hat{e}}_\theta }\\
 - \cos (m\phi ){{dP_l^m\left( {\cos \theta } \right)} \over {d\theta }}{\mathbf{\hat{e}}_\phi}.
\end{aligned}
\end{equation}
According to the previous analysis, the transformation above has flipped the helicity from $\sigma$ to $-\sigma$ within the rotating frame. As a result, the index for $\mathbb{M}^e_{lm}$ can be expressed as:
\begin{equation}
\label{index_pole_transform}
{\rm{\mathbb{I} \mathbb{N} \mathbb{D} (\theta=0,\pi)= }}1+|m|.
\end{equation}
As will be explained below, Eq.~(\ref{index_pole_transform}) is only valid for $m\neq \pm1$.\\

$\blacksquare$~~~\textsc{{Why a continuous $\mathbb{M}^e_{lm}$ Does not exist for $m=\pm 1$}}.\\

\begin{figure}[tp]
\centerline{\includegraphics[width=5cm]{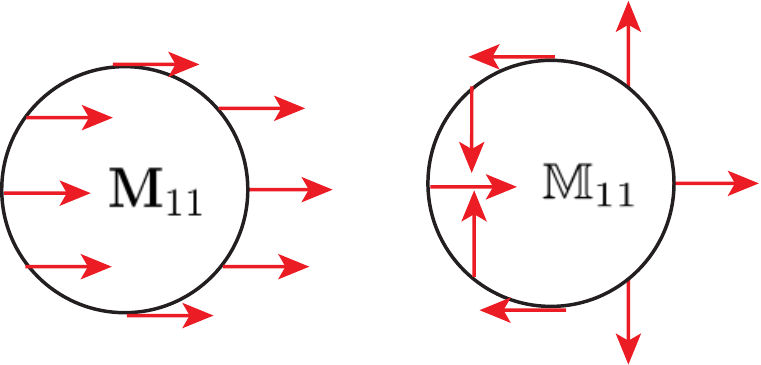}} \caption{\small Vector field patterns around the poles for both $\mathbf{M}^e_{l1}$ and $\mathbb{M}^e_{11}$. $\mathbf{M}^e_{l1}$ can be both non-vanishing and continuous at the poles, while $\mathbb{M}^e_{11}$  cannot be possibly both non-vanishing and continuous.}
\label{figs1}
\end{figure}

When $m= \pm 1$,  similar to $\mathbf{M}^e_{lm}$,  $|{{\mathbb{M}}^{e}_{lm}}|\neq0$ for $\theta=0$ or $\pi$. This means $\mathbb{M}^e_{l,\pm1}$ would be non-vanishing at the south or north poles. For a non-vanishing vector field at the poles, the continuity requires that all field vectors point to the same direction close to the poles. For $\mathbf{M}^e_{l,\pm1}$, this is exactly the case. Unfortunately this is not the case anymore for $\mathbb{M}^e_{l,\pm1}$, as according to our previous analysis, the field vectors close to pole would be exactly the same as that of a singularity of index $1+|m|$, rotating by an angle of $2(1+|m|)$ along the closed contour. For example, in Fig.~\ref{figs1} we show the vector field patterns around the poles for both $\mathbf{M}^e_{l1}$ and $\mathbb{M}^e_{11}$. It is clear that $\mathbf{M}^e_{11}$ can be both non-vanishing and continuous across the poles while this is not possible for $\mathbb{M}^e_{11}$.

\subsection{Singularity and index when $\theta\neq0$ or $\pi$}

In the regions excluding the poles, azimuthal angle $\phi$ is well defined and a singularity requires that both the both components along ${{\mathbf{\hat{e}}_\phi}}$ and  ${{\mathbf{\hat{e}}_\theta}}$ vanish. \\

$\blacksquare$~~~\textsc{Singularity and index for $\mathbf{M}^e_{lm}$}.\\

For $\mathbf{M}^e_{lm}$, a singularity requires that:
\begin{equation}
\label{nodal_crossing1}
\cos (m\phi)=P_l^m\left( {\cos \theta } \right)=0,
\end{equation}
or
\begin{equation}
\label{nodal_crossing2}
\sin (m\phi)=\frac{dP_l^m({\cos \theta })}{d\theta}=0.
\end{equation}
For both scenarios shown in  Eqs.~(\ref{nodal_crossing1}) and ~(\ref{nodal_crossing2}), the common feature is that  the singularity comes from the crossings of two nodal lines, which means that the index of the singularity would be either $+1$ (sink, source or center singularity) or $-1$ (saddle singularity). The exception of this is that when $m=0$: Eq.~(\ref{nodal_crossing1}) does not have solution; while for Eq.~(\ref{nodal_crossing2}), the nodal line of $\sin (m\phi)=0$ would then evolve into a nodal plane, for which we can assign an index of $0$. It is worth mentioning that for our study into multipolar radiations, center-type singularity are equivalent to sink (source)-type singularity, since if it is a sink (source) for electric field, then it becomes a center for magnetic field, and vice versa. \\

\begin{figure}[tp]
\centerline{\includegraphics[width=7cm]{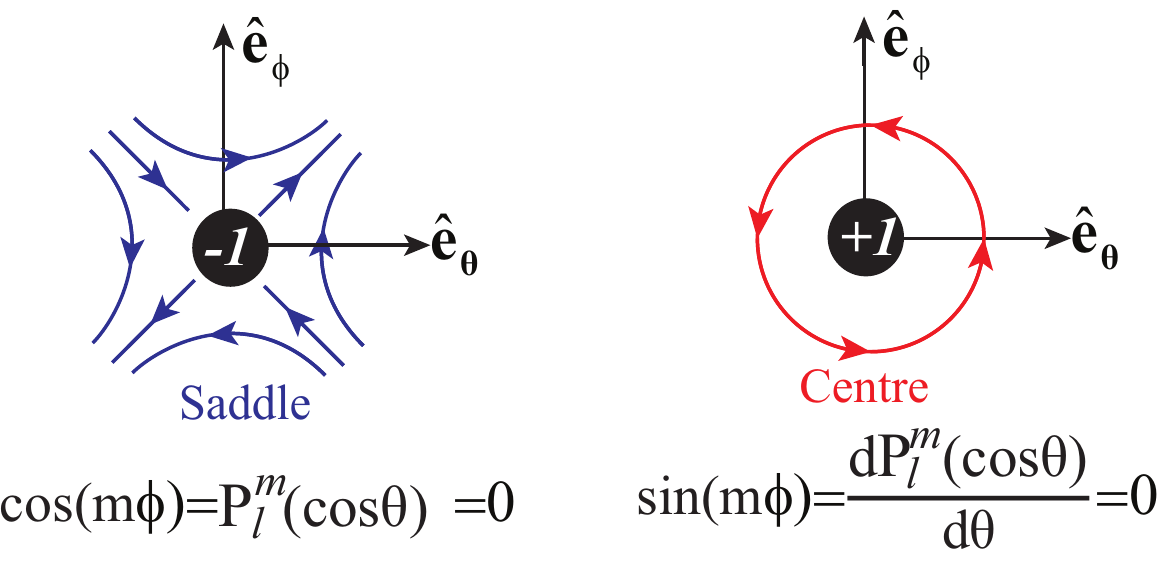}} \caption{\small Vector field patterns close to the singularities formed by the crossings of the two nodal lines described in Eqs.~(\ref{index_sm_1_1}) and (\ref{index_sm_1_2}). }
\label{figs3}
\end{figure}

For $0<\theta<\pi$ and $0\leq \phi \leq 2\pi$: both equations of $\sin (m\phi )=0$ and $\cos (m\phi )=0$ have $2|m|$ solutions ($m\neq0$); while equations of $P_l^m\left( {\cos \theta } \right)=0$ and $\frac{dP_l^m({\cos \theta })}{d\theta}=0$ have $l-|m|$ and $l-|m|+1$ solutions ($m\neq0$), respectively~\cite{BRONSHTEIN_2007__Handbook}. As a result, Eq.~(\ref{nodal_crossing1}) corresponds to $2|m|(l-|m|)$ singularities while Eq.~(\ref{nodal_crossing2}) corresponds to $2|m|(l-|m|+1)$ singularities. Both expressions are applicable for the special case of $m=0$, when there are no singularities located between the two poles. Rotational symmetry guarantees that the singularities described by the same equation share the same index of $-1$ or $+1$. When $m\neq0$, simple vectorial analysis of the two nodal lines (schematically shown in Fig.~\ref{figs3}) tells us that the singularities described in Eq.~(\ref{nodal_crossing1}) are of saddle-type and show index of $-1$; while those described in Eq.~(\ref{nodal_crossing2}) are of source (sink or centre)-type and thus show index of $+1$. This can be summarized as follows ($m\neq 0$):
\begin{eqnarray}
&{\mathbf{Ind}}=-1,~\cos (m\phi)=P_l^m\left( {\cos \theta } \right)=0; \label{index_sm_1_1}\\
&{\mathbf{Ind}}=+1,~\sin (m\phi)=\frac{dP_l^m({\cos \theta })}{d\theta}=0.\label{index_sm_1_2}
\end{eqnarray}
 Over the entire momentum sphere, the index sum for all singularities is: \\
\begin{equation}
\label{index_sum_bracket}
\underbrace {2(1 - \left| m \right|)}_{\rm{\mathbf{Poles}}}  \underbrace {+2m(l - \left| m \right| + 1)}_{\rm{\mathbf{Eq.(S13)}}}  \underbrace {-2m(l - \left| m \right|)}_{\rm{\mathbf{Eq.(S12)}}}
\end{equation}
which would always make up to $2$, agreeing with the Poincar\'{e}-Hopf theorem~\cite{MILNOR_1997__Topology,RICHESON_2012__Euler}.\\

$\blacksquare$~~~\textsc{Singularity and index for $\mathbb{M}^e_{lm}$}$(|m|\neq1)$.\\

\begin{table}[ht]
\centering 
\begin{tabular}{| c | c | c | c|} 
\hline 
Positions & Index & Number & Total Indexes \\ [0.5ex] 
\hline 
$|\cos(\theta)|=1$ & $1+|m|$ & 2 & $2(1+|m|)$ \\ 
\hline
$\cos (m\phi )=P_l^m\left( {\cos \theta } \right)$ & & $2|m|\times$ & $2|m|\times$\\$=0,|\cos(\theta)|\neq1$  & \raisebox{1.5ex}{+1 $(m\neq0)$} &  \raisebox{0ex}{$(l-|m|)$} & \raisebox{0ex}{$(l-|m|)$} \\
\hline 
\raisebox{-1.3ex}{$\sin (m\phi )=\frac{dP_l^m({\cos \theta })}{d\theta}$} & \raisebox{-0.75ex} {0 $(m=0)$} & \raisebox{-0.75ex} {$2|m|(l-$} & \raisebox{-0.75ex}{$-2|m|\times$}\\$=0,|\cos(\theta)|\neq1$  & \raisebox{0.75ex} {-1 $(m\neq0)$} &  \raisebox{0.75ex}{$|m|+1)$} & \raisebox{0.75ex}{$(l-|m|+1)$} \\
\hline
\end{tabular}\\
\caption{Distributions of the singularities and their indexes for $\mathbb{M}^e_{lm}(|m|\neq1)$.} 
\label{table_sm} 
\end{table}

During the transformation of $\mathbf{\hat{e}}_\theta \rightarrow -\mathbf{\hat{e}}_\theta$ (or equivalently $\mathbf{\hat{e}}_\phi \rightarrow -\mathbf{\hat{e}}_\phi$), the positions of the singularities would be fixed [see Eqs.~(\ref{nodal_crossing1}) and (\ref{nodal_crossing2})] while the sign of the index would be flipped: a saddle would be transformed to a sink (source) and vice versa. As a result, the singularity and index distributions for $\mathbb{M}^e_{lm}(|m|\neq1)$ is as follows ($m\neq0$):
\begin{eqnarray}
\label{index_sm_2}
&{\mathbb{IND}}=+1,~\cos (m\phi)=P_l^m\left( {\cos \theta } \right)=0;\\
&{\mathbb{IND}}=-1,~\sin (m\phi)=\frac{dP_l^m({\cos \theta })}{d\theta}=0.
\end{eqnarray}
For $\mathbb{M}^e_{lm}(|m|\neq1)$, over the entire momentum sphere, the index sum for all singularities is:
\begin{equation}
\label{index_sum_bracket2}
\underbrace {2(1 + \left| m \right|)}_{\rm{\mathbf{Poles}}}  \underbrace {-2m(l - \left| m \right| + 1)}_{\rm{\mathbf{Eq.(S16)}}}  \underbrace {+2m(l - \left| m \right|)}_{\rm{\mathbf{Eq.(S15)}}},
\end{equation}
which would also always make up to $2$, as is required by Poincar\'{e}-Hopf theorem~\cite{MILNOR_1997__Topology,RICHESON_2012__Euler}. The singularities and their index distributions for {$\mathbb{M}^e_{lm}$} $(|m|\neq1)$ has been summarized in Table ~\ref{table_sm}.

\section{Radiation patterns, singularities and indexes for $\mathbf{M}^e_{lm}$ with $l=1\rightarrow3$ and $m=0\rightarrow l$}

\begin{figure}[tp]
\centerline{\includegraphics[width=9cm]{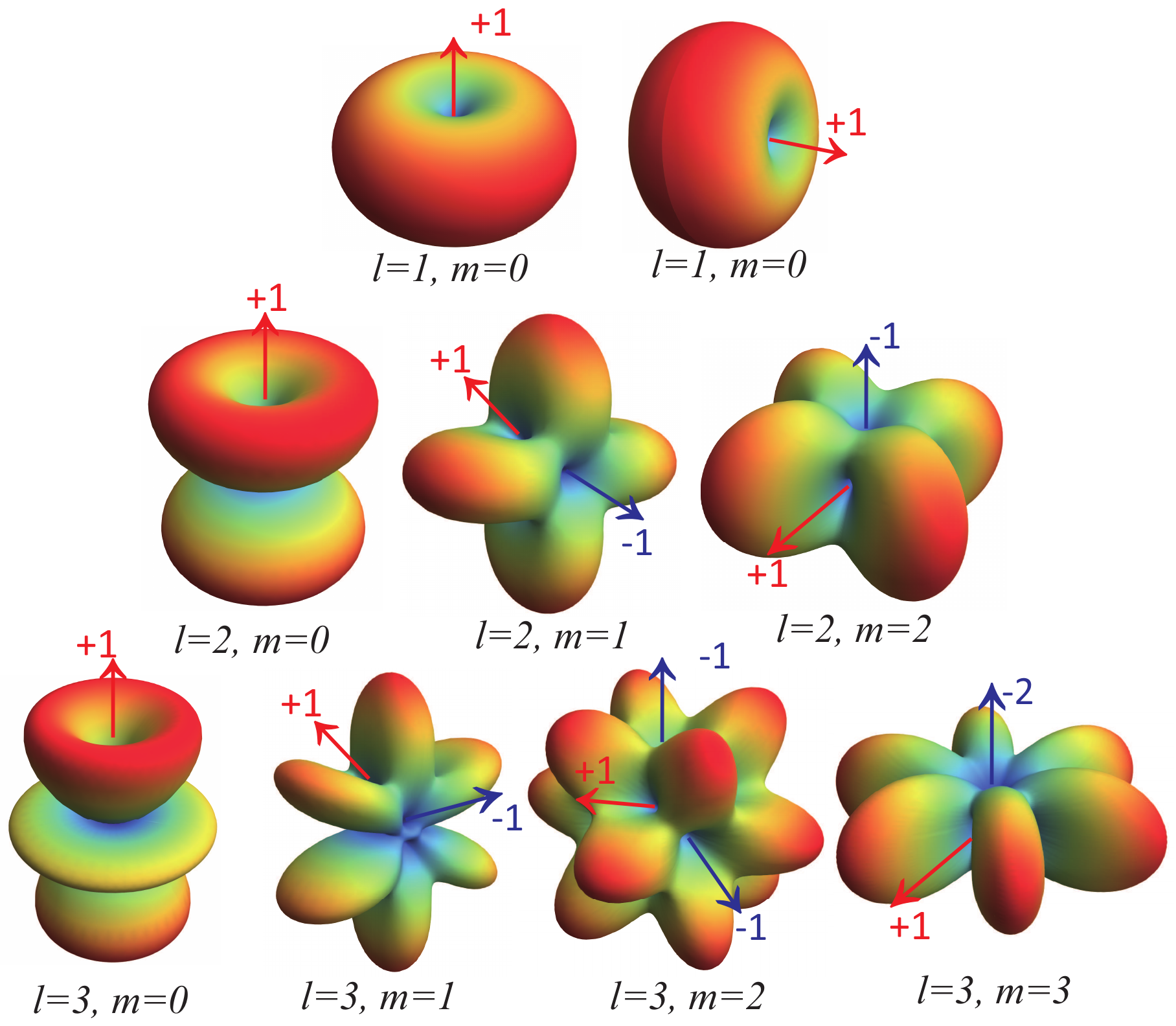}} \caption{\small Radiation patterns for multipoles of order $(l, m)$: $l=1\rightarrow3$ and $m=0\rightarrow l$. Representative singularities are pinpointed with their indexes specified.}
\label{figs2}
\end{figure}

Figure~\ref{figs2} shows the radiation patterns for multipoles described by $\mathbf{M}^e_{lm}$ with $l=1\rightarrow3$ and $m=0\rightarrow l$. Singularities that represent each category in Table I in the main letter are pinpointed and the corresponding indexes are specified.

\section{Multipolar expansions and the radiation patterns for unit cells of infinite periodic structures}

For finite particles or particles clusters, the radiated  fields along all directions can be (numerically) calculated, which can then be directly expanded to spherical harmonics of different orders with coefficients of $a_{lm}$ and $b_{lm}$. Then the magnitudes and phases of all electric and magnetic multipoles would be obtained, as has been finely implemented in most previous studies~\cite{jahani_alldielectric_2016,KUZNETSOV_Science_optically_2016,LIU_2018_Opt.Express_Generalized}. This approach of radiation expansion (expansion of the radiated fields) can not be directly implemented for unit cells of infinite periodic structures, as the radiation of infinite periodic structures is only known and allowed along certain directions of diffractions. \\

For infinite periodic structures, we can carry out the multipolar expansion based on the near-field polarization currents $\bf{J(r)}$ of the unit cell. The multipolar coefficients of $a_{lm}$ and $b_{lm}$ can be obtained through the following relations (when there are no effective magnetization currents)~\cite{jackson1962classical,GRAHN_NewJ.Phys._electromagnetic_2012}:
\begin{eqnarray}
&{a_{lm}} = {{{{\left( { - i} \right)}^{l - 1}}{k^2}\eta {Q_{lm}}} \over {{{\left[ {\pi \left( {2l + 1} \right)} \right]}^{{1 \over 2}}}}}\mathop{\int\!\!\!\int\!\!\!\int} \limits_\mathbf{V } {{\bf{J}}\left( {\bf{r}} \right) \cdot {\mathbf{S}_{lm}}{d^3}r}; \label{current_expansion_1}\\
&{b_{lm}} = {{{{\left( { - i} \right)}^{l + 1}}{k^2}\eta {Q_{lm}}} \over {{{\left[ {\pi \left( {2l + 1} \right)} \right]}^{{1 \over 2}}}}}\mathop{\int\!\!\!\int\!\!\!\int}\limits_\mathbf{V} {{\bf{J}}\left( {\bf{r}} \right) \cdot {\mathbf{T}_{lm}}{d^3}r},\label{current_expansion_2}
\end{eqnarray}  
where $\eta$ is the vacuum impendence; ${Q_{lm}} = {1 \over {{{\left[ {l\left( {l + 1} \right)} \right]}^{1/2}}}}{\left[ {{{2l + 1} \over {4\pi }}{{\left( {l - m} \right)!} \over {\left( {l + m} \right)!}}} \right]^{1/2}}$; the integration is conducted within the unit-cell region \textbf{V}. As long as the electric field distributions ${\bf{E}}\left( {\bf{r}} \right)$ within the unit cell is obtained, the corresponding electric polarzation currents can be obtained as:
\begin{equation}
\label{electric current}
{\bf{J}}\left( {\bf{r}} \right) =  - i\omega {\varepsilon _0}\left[ {{\varepsilon _r}\left( {\bf{r}} \right) - {\varepsilon _b}} \right]{\bf{E}}\left( {\bf{r}} \right),
\end{equation}
where ${\varepsilon _r}\left( {\bf{r}} \right)$ denotes the permittivity distribution and ${\varepsilon _b}$ is the permittivity of the background medium. Vector functions $\mathbf{S}_{lm}$ and $\mathbf{T}_{lm}$ are:
\begin{eqnarray*}
\label{vector_function}
\begin{split}
{\mathbf{S}_{lm}} = \exp \left( { - im\phi } \right)\left[ {{\Pi _l}\left( {kr} \right) + {\Pi _l}^{\prime \prime }\left( {kr} \right)} \right]P_l^m\left( {\cos \theta } \right){\hat{\mathbf{e}}_r} \\
+ \exp \left( { - im\phi } \right){{{\Pi _l}^\prime \left( {kr} \right)} \over {kr}}\left[ {{\tau _{lm}}\left( \theta  \right){\hat{\mathbf{e}}_\theta } - i{\pi _{lm}}\left( \theta  \right){\hat{\mathbf{e}}_\phi }} \right]
\end{split};\\
{\mathbf{T}_{lm}} = \exp \left( { - im\phi } \right){j_l}\left( {kr} \right)\left[ {i{\pi _{lm}}\left( \theta  \right)\hat{\mathbf{e}}_ \theta  + {\tau _{lm}}\left( \theta  \right)\hat{\mathbf{e}}_ \phi } \right],
\end{eqnarray*}
for which ${\tau _{lm}}\left( \theta  \right) = {d \over {d\theta }}P_l^m\left( {\cos \theta } \right)$, ${\pi _{lm}}\left( \theta  \right) = {m \over {\sin \theta }}P_l^m\left( {\cos \theta } \right)$, and ${\Pi _l}\left( {kr} \right) = kr{j_l}\left( {kr} \right)$ is Riccati-Bessel function [$j_l(kr)$ is the spherical Bessel function of the first kind]~\cite{BRONSHTEIN_2007__Handbook}.\\

\begin{figure}[tp]
\centerline{\includegraphics[width=8.8cm]{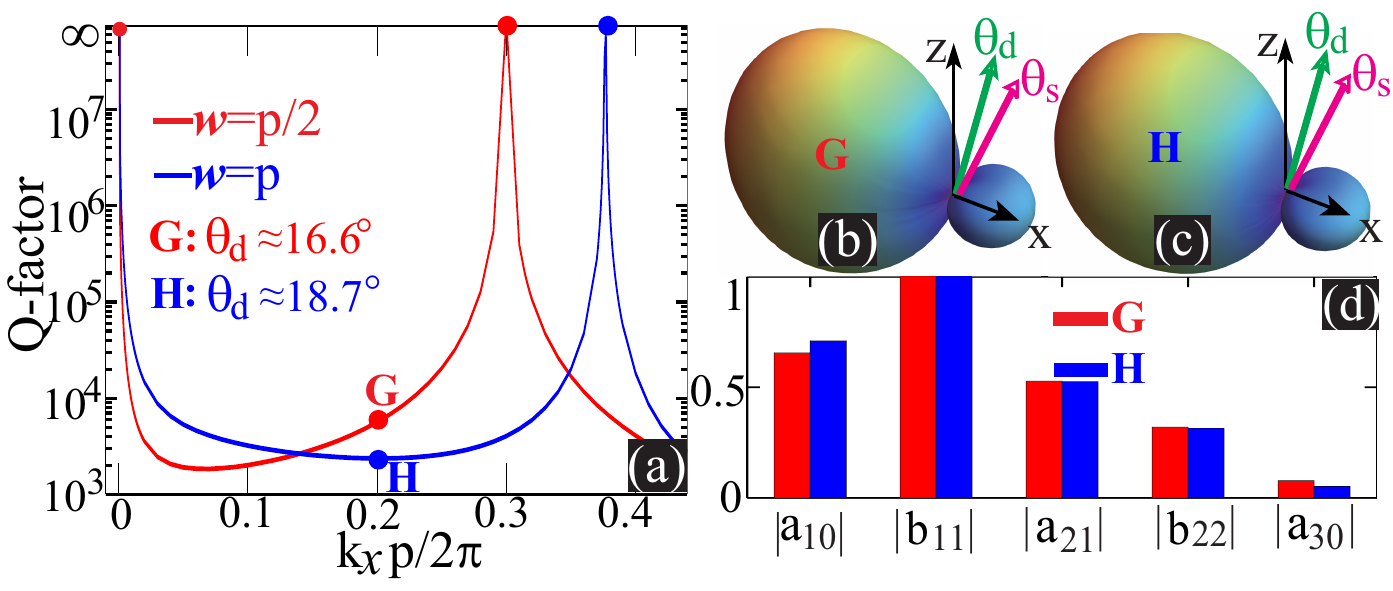}} \caption{\small (a) With width $w$ chaged from $p/2$ to $p$, two decay (non-BIC) states has been indicated by points \textbf{G} and \textbf{H} ($k_xp/2\pi=0.2$). The multipolar compositions for the two states are shown in (d) (only the major contributing terms are included), and the reconstructed far-field radiation patterns are shown in (b) and (c). For both cases the singularities (\textbf{G}: $\phi_s=0$ and $\theta_s\approx23.5^\circ$; \textbf{H}: $\phi_s=0$ and $\theta_s\approx26.5^\circ$) are of index $+1$  and do not overlap with the open diffraction channels (\textbf{G}: $\phi_d=0$ and $\theta_d\approx16.6^\circ$; \textbf{H}: $\phi_d=0$ and $\theta_d\approx18.7^\circ$).}
\label{figs4}
\end{figure}

With the expansion coefficients $a_{lm}$ and $b_{lm}$  obtained through Eqs.~(\ref{current_expansion_1}) and (\ref{current_expansion_2}), the radiated fields from the unit cell can be reconstructed as follows:
\begin{equation}
\label{radiated_fields}
{{\bf{E}}_{\rm{rad}}}\left( {r,\theta ,\phi } \right) = \sum\limits_{l = 1}^\infty  {\sum\limits_{m =  - l}^l {{E_{lm}}\left[ {{a_{lm}}{{\bf{\breve{N}}}_{lm}} + {b_{lm}}{{\bf{\breve{M}}}_{lm}}} \right]} },
\end{equation}
where ${E_{lm}} = {{{i^{l + 1}}\left( {2l + 1} \right)} \over 2}\sqrt {{{\left( {l - m} \right)!} \over {l\left( {l + 1} \right)\left( {l + m} \right)!}}}$; and vector functions $\bf{\breve{N}}$$_{lm}$ and $\bf{\breve{M}}$$_{lm}$ are:
\begin{eqnarray*}
\label{vector_function_radiation}
\begin{split}
{{{\bf{\breve{N}}}}_{lm}}{\rm{ = }}\left[ {{\tau _{lm}}\left( {\cos \theta } \right){{{\bf{\hat e}}}_\theta } + i{\pi _{lm}}\left( {\cos \theta } \right){{{\bf{\hat e}}}_\phi }} \right]{{{{\left[ {kr{h_l}^{\left( 1 \right)}\left( {kr} \right)} \right]}^\prime }} \over {kr}}\\
+ \exp \left( {im\phi } \right){{{\bf{\hat e}}}_r}l\left( {l + 1} \right)P_l^m\left( {\cos \theta } \right){{{h_l}^{\left( 1 \right)}\left( {kr} \right)} \over {kr}}\exp \left( {im\phi } \right);
\end{split}\\
{\mathbf{\breve{M}}_{lm}} = \left[ {i{\pi _{lm}}\left( {\cos \theta } \right){{{\bf{\hat e}}}_\theta } - {\tau _{lm}}\left( {\cos \theta } \right){{{\bf{\hat e}}}_\phi }} \right]{h_l}^{\left( 1 \right)}\left( {kr} \right)\exp \left( {im\phi } \right),
\end{eqnarray*}
for which ${h_l}^{\left( 1 \right)}\left( {kr} \right)$ is the spherical Hankel function of the first kind~\cite{BRONSHTEIN_2007__Handbook}.

\begin{figure}[tp]
\centerline{\includegraphics[width=7cm]{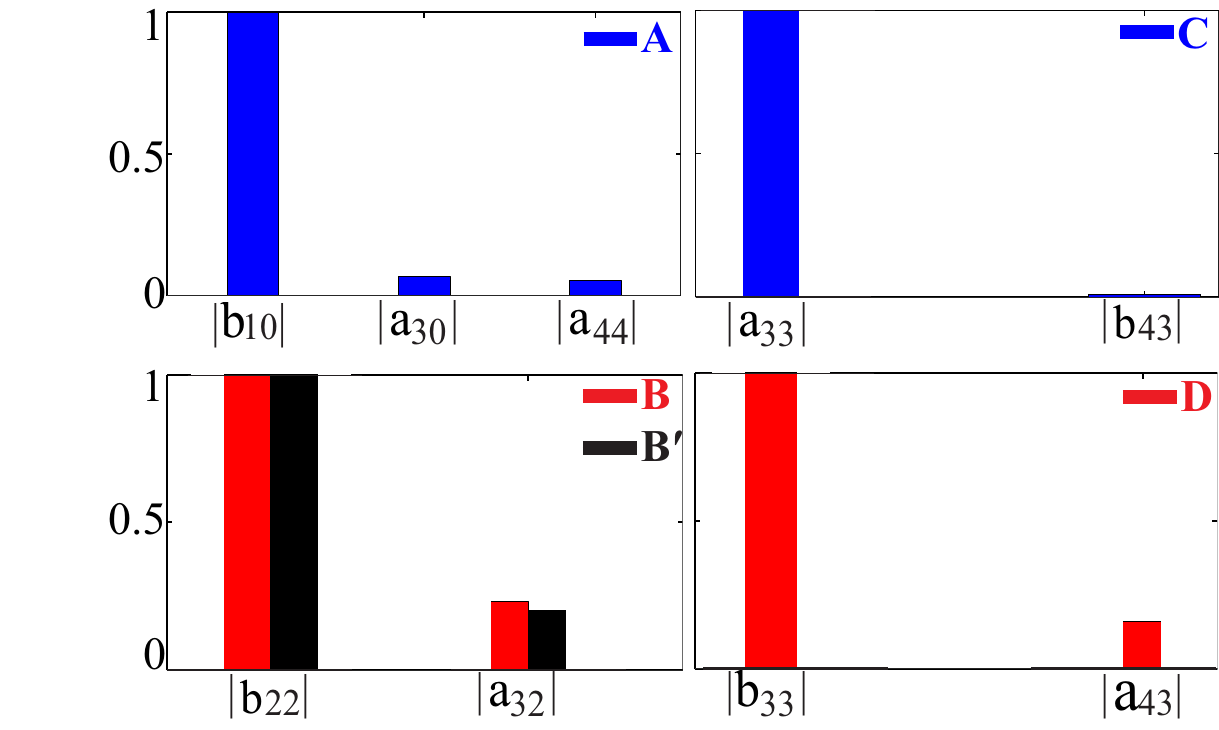}} \caption{\small Multipolar compositions (normalized magnitudes of expansion coefficients) for the $\Gamma$-point BICs indicated in the main letter by \textbf{A}-\textbf{D} and $\mathbf{B^\prime}$.}
\label{figs5}
\end{figure}

\section{Numerical calculations of the dispersion bands, the mode Q-factors and the field distributions within the unit cells of photonic crystal slabs}
For the photonic crystal slabs studied in this work, we employ a commercial software package COMSOL Multiphysics(https://www.comsol.com) to calculate the the dispersion bands. Each mode located on the band is characterized by complex eigenfrequencies with both  real and imaginary parts: $\widetilde\omega=\widetilde\omega_1+i\widetilde\omega_2$.  With $\widetilde\omega$  obtained, the Q-factor of the mode can be calculated as:
\begin{equation}
\label{q-fctor}
Q=\frac{\widetilde\omega_1}{2\widetilde\omega_2}.
\end{equation}
Both the complex eigenfrequencies and electric fields within the unit cell ${\bf{E}}\left( {\bf{r}} \right)$ that are employed to obtain the electric polarization currents ${\bf{J}}\left( {\bf{r}} \right)$ [see Eq.~(\ref{electric current})], can be calculated using COMSOL.

\section{Decay (non-BIC) states and separations of multipolar singularities from open radiation channels}

As is shown in the main letter, the formation of decay-free BICs can be attributed to the overlapping of multipolar singularity with the open radiation channel. When the singularity is separated from the open channel and radiation is not zero along the diffraction direction, energy leakage is allowed and then the BICs would evolve into decay states.  In Fig.~\ref{figs4}(a) [the same as Fig.~4(a) in the main letter] we have indicated two decay states by \textbf{G} ($\phi_d=0$ and $\theta_d\approx16.6^\circ$) and \textbf{H} ($\phi_d=0$ and $\theta_d\approx18.7^\circ$), and show the corresponding radiation patterns, singularities, indexes, diffraction directions and multipolar compositions in Figs.~\ref{figs4}(b)-(d). For both cases, the singularities (\textbf{G}: $\phi_s=0$ and $\theta_s\approx23.5^\circ$; \textbf{H}: $\phi_s=0$ and $\theta_s\approx26.5^\circ$) do not overlap with the open radiation channels to prevent energy leakage, and thus the indicated modes are not bound states, showing finite Q-factors.

\section{Multipolar compositions for the $\Gamma$-point BICs}
Specific multipolar compositions (normalized magnitudes of expansion coefficients) for the $\Gamma$-point BICs indicated in the main letter by \textbf{A}-\textbf{D} and $\mathbf{B^\prime}$ are summarized in Fig.~\ref{figs5}. For all cases, the multipoles of $|m|=1$ (the radiations of which are not zero at poles) do not exist and thus the radiations are intrinsically zero and singular along $\mathbf{z}$ direction.

\begin{figure}[tp]
\centerline{\includegraphics[width=8.5cm]{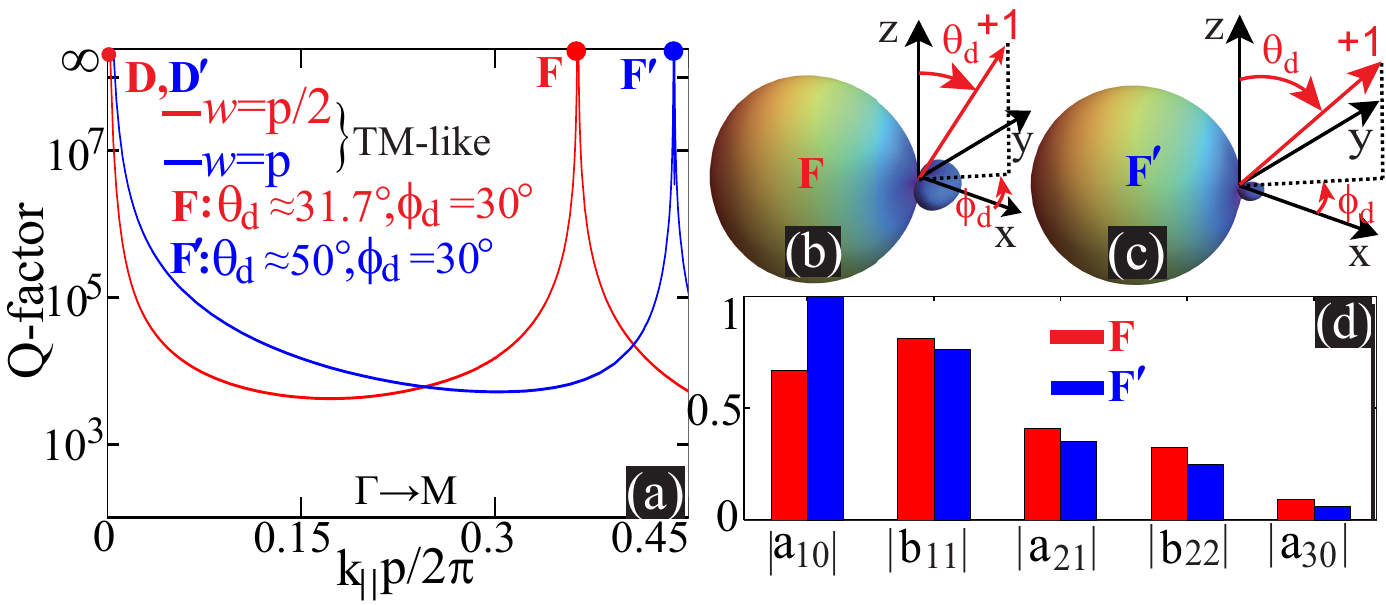}} \caption{\small (a) With width $w$ changed from $p/2$ to $p$, the off-$\Gamma$-point BIC on the TM-like branch of the hexagonal photonic crystal slab is shifted from $\textbf{H}$ ($k_\|p/2\pi=0.364$) to $\textbf{H$^\prime$}$ ($k_\|p/2\pi=0.438$), while the $\Gamma$-point BIC is fixed. The multipolar compositions for the two off-$\Gamma$-point BICs are shown in (d) (only the major contributing terms are included), and the reconstructed far-field radiation patterns are shown in (b) and (c). For both cases the singularity is of index $+1$  and overlaps with the open diffraction channels (\textbf{H}: $\phi_d=30^\circ$ and $\theta_d\approx31.7^\circ$; $\textbf{H$^\prime$}$: $\phi_d=30^\circ$ and $\theta_d\approx50^\circ$).}
\label{figs6}
\end{figure}

\section{Off-$\Gamma$-point BICs of the hexagonal photonic crystal slab}
Besides the square photonic crystal slab, off-$\Gamma$-point accidental BICs are also supported by the hexagonal photonic crystal slab, one of which is indicated by \textbf{F} ($k_\|p/2\pi=0.364$) in Fig.~\ref{figs6}(a) [geometric parameters are the same as those in Fig.~3(d) in the main letter and this indicated point is on the TM-like branch along $\Gamma$-M direction]. This bound state would moved to another position $\textbf{F$^\prime$}$ ($k_\|p/2\pi=0.438$) by changing $w$ from $p/2$ to $p$, as is shown in Fig.~\ref{figs6}(a). We further show respectively the multipolar compositions and reconstructed far-field patterns in Figs.~\ref{figs6}(b)-(d). Similar to the square photonic crystal slab, for both BICs the singularities of index +1  overlaps with the allowed diffraction directions (\textbf{F}: $\phi_d=30^\circ$ and $\theta_d\approx31.7^\circ$; $\textbf{F$^\prime$}$: $\phi_d=30^\circ$ and $\theta_d\approx50^\circ$), where $\phi_d=\arccos(k_{x}/k_{\|})$, $\theta_d=\arcsin(k_{\|}/k_0)$, and $k_0$ is total the angular wavenumber. This confirms our multipolar reinterpretation of the building mechanism of BICs and the connections between singularity indexes and their topological charges.

\end{document}